\documentclass{elsart}

\usepackage{graphicx}
\usepackage{dcolumn}
\usepackage{bm}
\usepackage{amsmath}

\newcommand{\ket}[1]{|#1\rangle}

\newcommand{\cn}{\frac{1}{\sqrt{2}}}

\begin{document}

\begin{frontmatter}
\title{Decoherence in the quantum walk on the line}
\author{A. Romanelli, }
\author{R. Siri, }
\author{G. Abal, }
\author{A. Auyuanet,}
\author{and R. Donangelo\thanksref{UFRJ}}
\address{Instituto de F\'{\i}sica, Facultad de Ingenier\'{\i}a\\
Universidad de la Rep\'ublica\\ C.C. 30, C.P. 11000, Montevideo, Uruguay}
\thanks[UFRJ]{Corresponding Author. Permanent address: Instituto de F\'{\i}sica,\\ 
Universidade Federal do Rio de Janeiro\\
C.P. 68528, 21941-972 Rio de Janeiro, Brazil}
\date{\today}
\begin{abstract}
We investigate the quantum walk on the line when decoherences are introduced
either through simultaneous measurements of the chirality and particle position,
or as a result of broken links.
Both mechanisms drive the system to a classical diffusive behavior. 
In the case of measurements, we show that the diffusion coefficient is proportional to the variance of the initially localized quantum random 
walker just before the first measurement. 
When links between neighboring sites are randomly broken with probability
$p$ per unit time, the evolution becomes decoherent after a characteristic
time that scales as $1/p$. The fact that the quadratic increase of the 
variance is eventually lost even for very small frequencies of disrupting
events, suggests that the implementation of a quantum walk on a real physical 
system may be severely limited by thermal noise and lattice imperfections.
\end{abstract}
\begin{keyword}
 Hadamard walk; quantum information; random walk; Markov process; Brownian motion
\end{keyword}
\end{frontmatter}

\section{Introduction}

One of the most challenging problems in quantum computation has been the design
of quantum algorithms which outperform their classical counterparts in meaningful
tasks. Few quantum codes in this category have been discovered after the well
known examples by Shor and Grover \cite{Shor,Grover}. Since the classical codes
based on the random walk process have been extremely well succeeded in certain tasks
\cite{clas_alg}, the hope that the quantum random walk \cite{hadamard} may provide a
similar insight for quantum coding has generated a great interest in this model.
However, the dynamical properties of the quantum random walk are rich enough so that
this system is physically interesting in its own right \cite{Kempe}. For instance,
due to quantum coherence effects, the position distribution of the quantum random
walker spreads out linearly in time. In contrast, the classical random walker
spreads out only as the square root of time.

Several systems have been proposed as candidates to implement quantum 
random walks. These proposals include atoms trapped in optical lattices
\cite{Dur}, cavity quantum electrodynamics (CQED) \cite{Sanders} and 
nuclear magnetic resonance (NMR) in solid substrates \cite{Du,Berman}. 
In liquid-state NMR systems \cite{NMR}, time--resolved observations of spin 
waves has been done \cite{NMR-exp}.
It has also been pointed out that a quantum walk can be simulated using 
classical waves instead of matter waves \cite{Roldan,Jeong}.

All these proposed implementations face the obstacle of decoherence due to environmental noise
and imperfections. Decoherence in the quantum walk on the line has been
considered recently by several authors. Numerical simulations of the effect
of different kinds of measurements have shown that the quantum walk properties are highly sensitive
to decoherent events and in particular that the quadratic increase of the variance is
eventually suppressed \cite{Kendon}. Other studies focused on the effect 
of measurements in chirality and reached similar conclusions \cite{Brun,Lopez}. 
A different decoherent mechanism, unitary noise, also leads to a crossover 
from a quantum behavior at short times to a classical-like behavior at 
longer times \cite{Shapira}. It is clear that the quadratic increase in 
the variance of the quantum walk with time is a direct consequence of the
coherence of the quantum evolution \cite{hadamard}. This can also
be visualized through the separation of the evolution equation for a
quantum system into a Markovian term and a quantum interference term as 
proposed in \cite{master}. In this work, besides the study of the effect
of measurements on the evolution of quantum random walkers, we also consider 
the decoherences generated by a different process, namely the influence of 
randomly broken links on the dynamics of the quantum walk. 
These mechanisms may be relevant in experimental realizations of quantum 
computers based on Ising spin--1/2 chains in solid-state substrates \cite{Berman}.

The paper is organized as follows. In the next section, we briefly introduce
the basic notions and notation relative to the discrete--time quantum walk
on the line. Then, in section~\ref{sec:medidas}, we consider joint periodic
measurements of chirality and position and show how the resulting
evolution can be described in terms of a master equation. 
In section~\ref{sec:blinks}
decoherence is introduced through random failures in the links between
neighboring sites. In section~\ref{sec:Brownian}, we show how the
stochastic classical model of Brownian motion can be used to describe the time dependence of the variance of the decoherent quantum walk.
Finally, in section~\ref{sec:conclusions}, we summarize our conclusions.

\section{Quantum random walk on the line}

Let us consider a particle that can move freely over a series of
interconnected sites. The discrete quantum walk on the line may be
implemented by introducing an additional degree of freedom, the chirality,
which can take two values: ``left'' or ``right'', $|L\rangle$ or $|R\rangle$,
respectively. This is the quantum analog of the coin--flipping decision
procedure for the classical random walker. At every time step, a rotation
(or, more generally, a unitary transformation) of the chirality takes place
and the particle moves according to its final chirality state. The global
Hilbert space of the system is the tensor product $H_{s}\otimes H_{c}$ where
$H_{s}$ is the Hilbert subspace associated to the motion on the line and the
chirality Hilbert subspace is $H_{c}$.

If one is only interested in the properties of the probability distribution,
it suffices to consider unitary transformations which can be expressed
in terms of a single real angular parameter $\theta$ \cite{Nayak,Tregenna,Bach}.
Let us call the operators that translate the walker one site to the left (right)
on the line in $H_{s}$ as $T_{L}$ ($T_{R} $), respectively, and let
$|L\rangle\langle L|$ and $|R\rangle\langle R|$ be the chirality projector
operators in $H_{c}$. We consider transformations of the form \cite{Nayak},
\begin{equation}
U(\theta)=\left\{ T_{L}\otimes|L\rangle\langle L|+T_{R}\otimes|R\rangle
\langle R|\right\} \circ\left\{ I\otimes K(\theta)\right\} ,  \label{Ugen}
\end{equation}
where $K(\theta)=\sigma_{z}e^{i\theta\sigma_{y}}$ is an unitary operator
acting on $H_{c}$, $\sigma_{y}$ and $\sigma_{z}$ being the Pauli
matrices, and $I$ is the identity operator in $H_{s}$. The unitary operator
$U(\theta)$ evolves the state $|\Psi(t)\rangle$ by one time step,
\begin{equation}
|\Psi(t+1)\rangle=U(\theta)|\Psi(t)\rangle.  \label{evol1}
\end{equation}
The wavevector $|\Psi(t)\rangle$ is expressed as the spinor
\begin{equation}
\ket{\Psi(t)} = \sum\limits_{n=-\infty}^{\infty}
\left[
\begin{array}{c}
a_{n}(t)\\b_{n}(t)
\end{array}
\right]\ket{n},  \label{spinor}
\end{equation}
where the qubit $(a,b)^\mathbf{T}$ has an upper (lower) component associated to the left (right)
chirality and the states $\ket{n}$ are eigenstates of the position operator corresponding
to the site $n$ on the line. In this work, for simplicity we set $\theta=\pi/4$ in eq.~(\ref{Ugen})
so that the unitary operation on chirality reduces to a Hadamard operation,
\begin{equation}
K=\cn\left(
\begin{array}{cc}
1&1\\
1&-1
\end{array}
\right).\label{hadamard}
\end{equation}
The evolution corresponding to eq.~(\ref{evol1}), can then be written as the map
\begin{align}
a_{n}(t+1) & =\cn\left[a_{n+1}(t)+b_{n+1}(t)\right]  \label{map} \\
b_{n}(t+1) & =\cn\left[a_{n-1}(t)-b_{n-1}(t)\right].  \notag
\end{align}

The probability distribution for the walker position at time $t$ is given by
\begin{equation}
P_{n}(t)=|a_{n}(t)|^{2}+|b_{n}(t)|^{2}.  \label{prob}
\end{equation}
Note that the unitary evolution described by eq.~(\ref{map}) has the property that
after each time step two new sites are occupied. Therefore, if initialized
at site $0$, after $t$ steps the wavefunction is non-zero, at most, in the
$2t+1$ sites around the origin. From the map in eq.~(\ref{map}) it may be shown 
that the Hadamard walk has a characteristic interference profile with two 
wavefronts which propagate away from the origin with constant speed
$1/\sqrt{2}$ \cite{Kempe}. In what follows we consider the effect of performing measurements on the evolution of this system.

\section{Periodic Measurements}
\label{sec:medidas}

Let us take as an initial condition the random walker starting from the 
position eigenstate $\ket{0}$ with chirality
$\cn (1,i)^\mathbf{T}$. This choice results in a symmetric evolution with $P_n(t)=P_{-n}(t)$ \cite{Konno}.
The position and chirality of the walker are jointly measured every $T$ steps. Among the several alternatives for measuring chirality, we choose to measure 
it in such a way that the chirality is projected on the 
$\overset{\rightarrow }{y}$ direction by the $\sigma_y$ Pauli operator. 
Then, the qubit states $\cn (1,i)^\mathbf{T}$ and $\cn (1,-i)^\mathbf{T}$ 
are eigenstates of the measurement operator and the symmetric character of 
the evolution is preserved.

The wavefunction evolves according to the unitary map~(\ref{map}) for the
first $T$ steps. At time $t=T$ position and chirality are measured for the 
first time.
The probability that the result of the position measurement is $n$,
\textit{i.e.} that the wavefunction collapses into the eigenstate $\ket{n}$
of position, is
\begin{equation}
q_{n}\equiv P_{n}(T)\,.  \label{position}
\end{equation}
The probability distribution $\{q_{n}\}$ depends on the initial qubit state.
However, for the chirality measurement procedure described above, this distribution repeats itself after each measurement, although centered at different position eigenstates, as can be seen from (\ref{map}).

\begin{figure}[tbp]
\begin{center}
\includegraphics[scale=0.7]{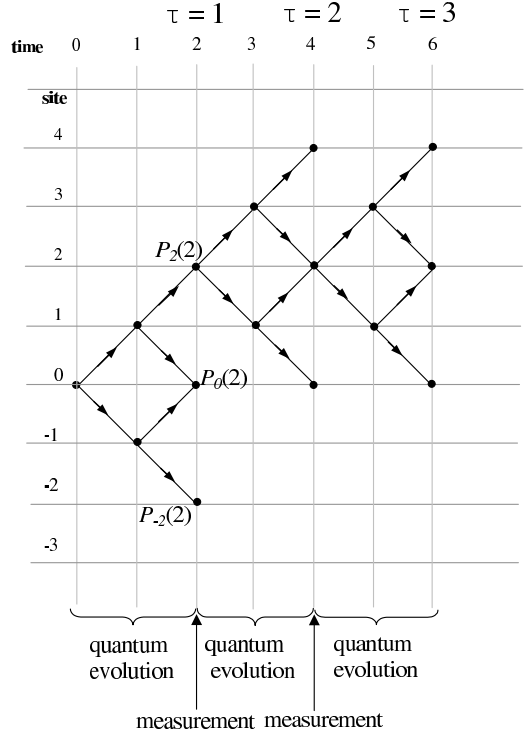}
\end{center}
\caption{{\protect Time evolution diagram for the quantum
random walker assuming measurements with a period of $T=2$ time steps. The
evolution shown assumes that the initial state in position of the system is
$n=0$.}}
\label{fig:dia1}
\end{figure}

After this first measurement, the unitary evolution continues and new
measurements are performed at times $t=2T,3T,\ldots \tau T$.
The discrete variable $\tau$ indicates the number of measurements performed 
on the system up to a given time. 
To illustrate this procedure, in Figure~\ref{fig:dia1}
we consider an example for the evolution in the case $T=2$.
For arbitrary intervals between consecutive measurements $T$, the probability
distribution $P_{n}$ satisfies the master equation
\begin{equation}
P_{n}(t + T)=\sum\limits_{j=n-T}^{n+T}q_{n-j}P_{j}(t),
\label{markov1}
\end{equation}
where the Markovian transition probabilities from site $j$ to site $n$,
$q_{n-j}$ are defined in eq.~(\ref{position}).

We calculate now, using eq.~(\ref{markov1}), the first moment as
$M_{1}(t)\equiv \sum jP_{j}(t)$ and the second moment as
$M_{2}(t)\equiv \sum j^{2}P_{j}(t)$. After some manipulation we have
\begin{align}
M_{1}(t + T)& =M_{1}(t)+M_{1q}(T)  \label{mom1} \\
M_{2}(t + T)& =M_{2}(t)+2M_{1}(t)M_{1q}(T)+M_{2q}(T)  \label{mom2}
\end{align}
where $M_{1q}(T)=\sum_{n=-T}^{n=T}nq_{n}$ and
$M_{2q}(T)=\sum_{n=-T}^{n=T}n^{2}q_{n}$
are the first and second moments associated to the unitary evolution between measurements. Therefore the variance,
$\sigma ^{2}(t)=M_{2}(t)-M_{1}^{2}(t)$, verifies
\begin{equation}
\sigma ^{2}(t +T) =\sigma ^{2}(t) +\sigma_{q}^{2}(T), \label{varia}
\end{equation}
where $\sigma _{q}^{2}(T)=M_{2q}(T)-M_{1q}^{2}(T)$ is the variance
associated to the unitary evolution between measurements.
\begin{figure}[tbp]
\begin{center}
\includegraphics[scale=0.5,angle=0]{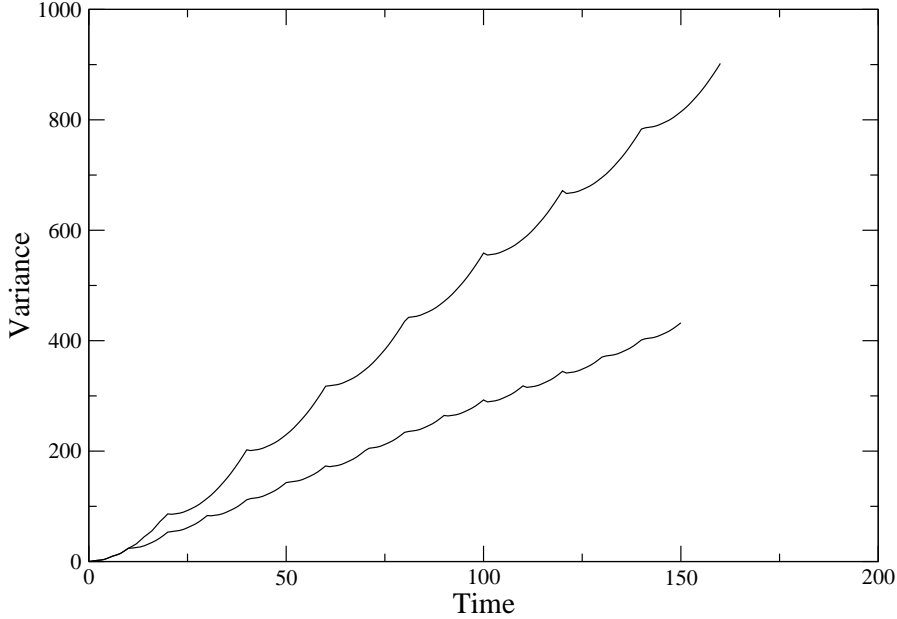} \end{center}
\caption{ Time evolution for the variance of the
quantum random walker. Two measurement rates are shown, corresponding to $T=10$ and $T=20$.
The parabolic segments correspond to the free quantum evolution between measurements.
The slope just before a measurement is proportional to the diffusion coefficient according
to eq.~(\ref{Dqrm0}). The initial condition is as mentioned in the text.}
\label{fig:dia2}
\end{figure}

From eq.~(\ref{varia}) the diffusion coefficient in the case of repeated
measurements, $D_{rm}=\frac12\frac{\Delta\sigma^2}{\Delta t}$ over a time scale
$\Delta t=\tau T$ including several measurement processes, is
\begin{equation}
D_{rm}=\frac{\sigma_{q}^{2}(T)}{2T}.
\label{Dqrm0}
\end{equation}
Since the variance of the free evolution of the quantum system increases
quadratically with time, we have
\begin{equation}
\sigma_{q}^{2}(T)=CT^{2}\,.  \label{sigmaq}
\end{equation}
Where, for $T\gg 1$, $C$ is a constant determined by the initial conditions.
Then, the diffusion coefficient depends linearly on the time interval between measurements,
\begin{equation}
D_{rm}=\frac{1}{2}CT.  \label{Drm}
\end{equation}
{\it i.e.} inversely with the frequency $f=1/T$ of decoherent events.

We show in Fig.~\ref{fig:dia2} the variance $\sigma ^{2}(t)$ calculated
through a computer simulation of the time evolution of an ensemble of
$10^4$ trajectories. These calculations confirm the linear dependence of
the diffusion coefficient with the time interval $T$ in eq.~(\ref{Drm}).
\section{Broken links}
\label{sec:blinks}

\begin{figure}[b]
\begin{center}
\includegraphics{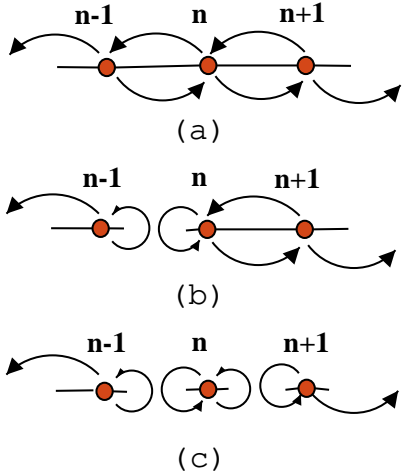}
\end{center}
\caption{  Situations that can arise at a site $n$ of the line
when there are: (a) no broken links, (b) the link to the left of the site
is broken, and (c) both links are broken. The upper (lower) arrows indicate
the direction of the probability flux associated to the upper (lower) spinor
component.}
\label{fig:lnk}
\end{figure}

Let us now consider a different mechanism for introducing decoherence in the quantum walk. 
Suppose that, at time $t$, a given site $n$ has one or both of the links connecting it to its neighboring sites broken.
If site $n$ has no broken links, as in  Figure~\ref{fig:lnk} (a), the
evolution law  (\ref{map}), which implies a Hadamard operation
in chirality space followed by a conditional translation (upper spinor
component to the left and the lower component to the right) is applied.
When one or both links at site $n$ are opened there can be no translation
across the broken link and the evolution must be modified accordingly.
If the link to the left of site $n$ is broken, as shown in
Fig.~\ref{fig:lnk} (b). The upper component of the spinor at $n$ receives
probability flux from $n+1$. In order to conserve flux, the outgoing probability
flux from the upper component at $n$ must be diverted to the lower component at
the same site and the corresponding transformation on the spinor components is,
\begin{eqnarray}
a_n(t+1)&=&\cn\left[a_{n+1}(t)+b_{n+1}(t)\right]\nonumber\\
b_n(t+1)&=&\cn\left[a_{n}(t)+b_{n}(t)\right].\label{left_blink}
\end{eqnarray}

If the broken link is to the right of $n$, the situation is similar, and the
transformation is
\begin{eqnarray}
	a_n(t+1)&=&\cn\left[a_{n}(t)-b_{n}(t)\right]\nonumber\\
	b_n(t+1)&=&\cn\left[a_{n-1}(t)-b_{n-1}(t)\right].
\label{right_blink}
\end{eqnarray}

Finally, if site $n$ is isolated, as in Fig.~\ref{fig:lnk} (c), the Hadamard
operation is followed by a chirality exchange,

\begin{eqnarray}
	a_n(t+1)&=&\cn\left[a_{n}(t)-b_{n}(t)\right]\nonumber\\
	b_n(t+1)&=&\cn\left[a_{n}(t)+b_{n}(t)\right].
\label{2blink}
\end{eqnarray}

The evolution consists of a sequence of unitary transformations applied
to the initial wavefunction
\begin{equation}
\ket{\Psi(t)}=U_tU_{t-1}\ldots U_1\ket{\Psi(0)}\label{unitary}
\end{equation}
and the form of each $U_k$ depends on the (random) topology of the line at
time step $t=k$. It follows that eqs.~(\ref{map}) and (\ref{left_blink})--(\ref{2blink}) preserve
the norm of the wavefunction, $\sum P_n(t) =1$, in the presence of an arbitrary
number of broken links.
In this sense, this model of decoherence is similar to the case of random unitary noise
discussed in \cite{Shapira} and is essentially different from repeated measurements,
which of course cannot be described as a sequence of unitary transformations.

We choose the same initial condition as in the previous section,
namely \\ $\ket{\Psi(0)}~=~\cn (1,i)^\mathbf{T}\ket{0}$, which
leads to a symmetric evolution in the coherent case. The algorithm for implementing
the quantum walk with broken links proceeds as follows. At each time step $t$, the state of
the links in the line is defined. Each link has a probability $p$ of breaking
in a given time step, $p$ being the only parameter in the model and the
Hadamard walk is recovered if $p=0$. Then, the corresponding transformation,
either of eqs.~(\ref{map}), (\ref{left_blink}), (\ref{right_blink}),
or (\ref{2blink}), is applied at each site.

We focused on two quantities: the probability distribution, eq.~(\ref{prob}),
at a fixed time and the time evolution
of its variance. As a result of decoherence, the distributions for $p\ne 0$ are
qualitatively different from the distribution for the unitary Hadamard walk.
The probability distributions for $p \ne 0$ approach a Gaussian shape in the
long time limit. At fixed time, the width of the Gaussian becomes narrower as
the probability of breaking links increases. For values of $p$ larger than about
$1/2$ the high frequency of broken links prevents the spread of the wavefunction 
beyond a certain region around the origin.
\begin{figure}
\centerline{\includegraphics[scale=0.75,angle=-90]{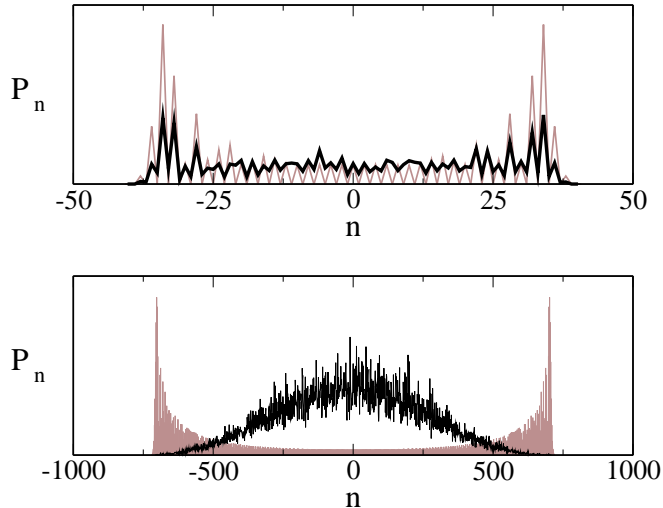}}
\caption{ Distribution $P_n$, from eq.~(\ref{prob}),
for $p=0.01$ at two different times. The upper panel corresponds to $t=50$
and the lower panel to $t=1000$. The corresponding distributions for the
case $p=0$ are shown in the background.}
\label{fig:distav}
\end{figure}

In order to find out how does the transition between the coherent quantum walk 
distribution and the Gaussian--like distributions take place we investigate the
low probability region of the model. In  Fig.~\ref{fig:distav}, the average profile 
$P_n$ for $p=0.01$ is shown at different times. At early times (upper panel) the 
distribution is still close to the coherent one while at long times (lower panel) 
the characteristic Gaussian profile has emerged. The corresponding profiles for 
the case $p=0$, are shown in the background as a reference. As disruptive events 
(in this model, the breaking of the links) take place, the distribution gradually 
shifts from the typical shape for the coherent quantum random walker to the one
characteristic of a classical walker.

There is a characteristic time, $t_c$, associated to this transition. A simple
argument can be used to understand the dependence of $t_c$ on the frequency $p$.
At early times, $t\ll t_c$, there are no disruptive events under the initially localized
wavefunction and it spreads coherently in both directions with speed $1/\sqrt{2}$,
covering a range $\sqrt{2}t$ in time $t$. The mean number of broken links per time step
under the wave function increases as $p\sqrt{2}t$. When this number becomes
of order 1, the effect of the disruptive events becomes relevant. This happens
at a coherence time
\begin{equation}
t_c=\frac{1}{p\sqrt{2}}.\label{tcoherence}
\end{equation}

\begin{figure}
\centerline{\includegraphics[scale=0.9,angle=-90]{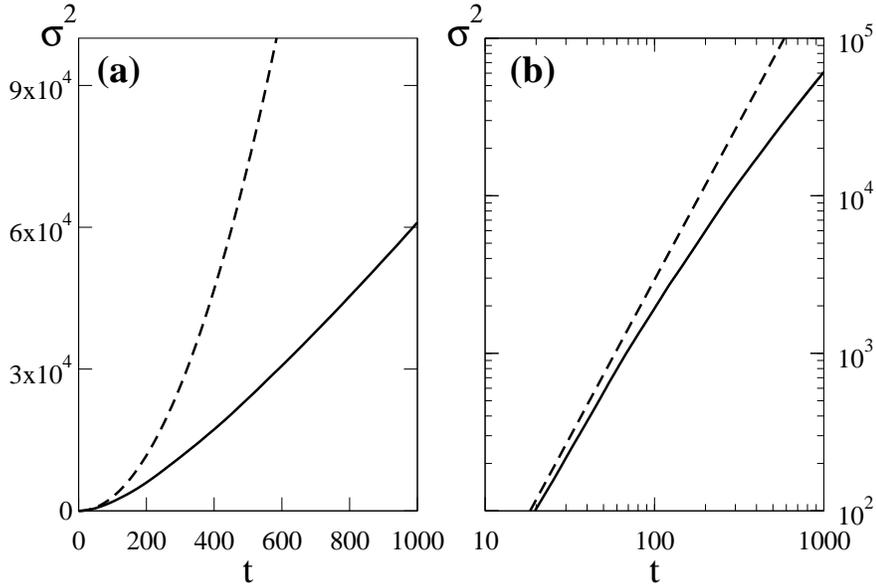}}
\caption{ (a) Time evolution of the variance, $\sigma^2$, for
$p=0.01$; in (b) the same evolution is shown in a log--log scale to emphasize that the transition
between a quadratic and a linear increase takes place gradually.
The dashed lines correspond to the evolution of the variance in the coherent case ($p=0$) which is
shown as a reference.}
\label{fig:mdv001}
\end{figure}

As mentioned before, for longer times $t\gg t_c$, the probability profile approaches
a Gaussian distribution. The transition is reflected in the variance of the distribution
as well. For early times, $t\ll t_c$, the variance shows the quadratic increase
characteristic of the unitary quantum walk. However, for non--zero $p$, a transition to
a linear increase regime takes place. Fig.~\ref{fig:mdv001} shows the time evolution of
$\sigma^2$ for $p=0.01$, for which eq.~(\ref{tcoherence}) implies a coherence time
$t_c\sim 70$. It is clear that the transition takes place gradually in time.

In the case of broken links, disruptive events take place at each time step so
the diffusion coefficient, $D_{bl}=\frac12\lim_{t\rightarrow\infty}\partial\sigma^2/\partial t$, 

\begin{figure}
\centerline{\includegraphics[scale=0.5,angle=-90]{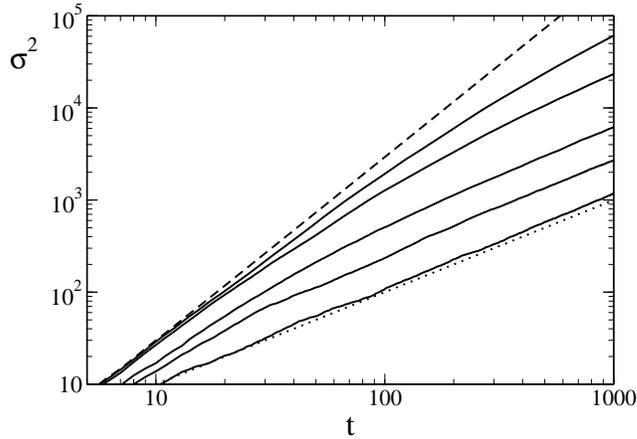}}
\caption{ Time evolution of the average variance of the quantum walker with broken
links in a log-log scale. The dashed line corresponds to the case $p=0$, and shows a quadratic
increase with time. The full lines correspond to several values of $p$: 0.01, 0.03, 0.10, 0.20 and 0.40. The dotted line corresponds to a classical unbiased random walk. }
\label{fig:mdv2av}
\end{figure}

\begin{figure}[b]
\centerline{\includegraphics[scale=0.6,angle=-90]{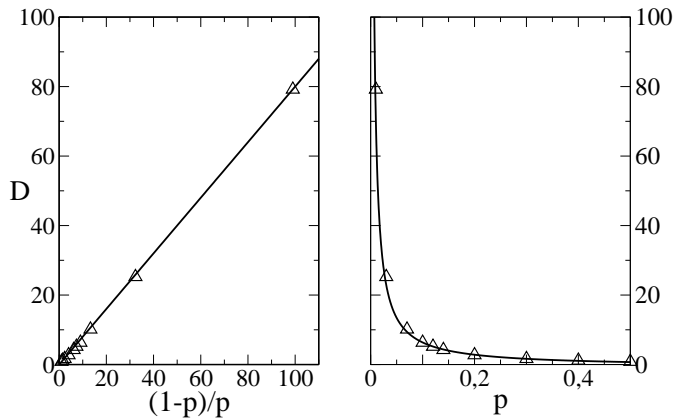}}
\caption{ Diffusion coefficient expressed as a function of $(1-p)/p$ (left panel)
and $p$ (right panel). The triangles are obtained from the data presented in
Fig.~\ref{fig:mdv2av}. The line corresponds to the linear fit to these data, eq.~(\ref{Dbl}).}
\label{fig:kpfit}
\end{figure}

In Fig.~\ref{fig:mdv2av}, the average time evolution of $\sigma^2$ for several $p$
values is shown in a log-log scale. From these results, an estimate for the diffusion coefficient can be obtained as a function of $p$. The results, shown in Fig.~\ref{fig:kpfit},
indicate a linear dependence with $(1-p)/p$,
\begin{equation}
D_{bl} = K\frac{1-p}{p}.\label{Dbl}
\end{equation}

A linear regression gives $K\simeq 0.40$ and this implies that for $p=4/9\sim 0,44$ the diffusion
coefficient is $1/2$ as in the unbiased classical random walk. For higher values of $p$, links are broken too frequently and the wavefunction is confined to a region about the origin.  this prevents full diffusion. For smaller link breaking frequencies the diffusion rates are higher than the classical rate, suggesting that correlations persist, as we show in Section~\ref{sec:Brownian}. This fact has been observed in the context of decoherence induced by non-unitary noise \cite{Brun}. Therefore, it should be kept in mind that the estimate for $K$, based on finite--time
data, is not exact because of residual correlations. As a consequence, it depends (weakly) on
the initial conditions. In the next section, where we consider the decoherent quantum walk from the
perspective of the Brownian motion model, we provide a simple analytical argument showing that the
assumption of complete correlation decay leads to incorrect results.

There is an interesting similarity between eqs~(\ref{Drm}) and (\ref{Dbl})
for the dependence of the diffusion coefficients in the case of repeated measurements and in the  evolution with broken links. In both cases, the dominant contribution to the diffusion coefficient depends inversely on the frequency of disruptive events. A similar dependence appears in other decoherent quantum walks \cite{Brun,Shapira}. This dependence is due to the persistence of quantum correlations.  
In order to investigate the role of quantum correlations in the diffusion coefficient, and in particular to see  how the corresponding classical diffusion scales with $p$, we consider the quantum evolution equations, eqs.~(\ref{map}) and (\ref{left_blink}--\ref{2blink}), written in terms of the occupation probabilities defined in (\ref{prob}),
\begin{eqnarray}
P_n(t+1)&=&\frac12\left[P_{n+1}(t)+P_{n-1}(t)\right]+\beta_{n+1}(t)-\beta_{n-1}(t)\nonumber\\
P_n(t+1)&=&\frac12\left[P_{n-1}(t)+P_{n}(t)\right]-\left[\beta_{n-1}(t)+\beta_{n}(t)\right]\label{Pev}\\P_{n}(t+1)&=&\frac12\left[P_{n}(t)+P_{n+1}(t)\right] + \beta_{n}(t)+\beta_{n+1}(t)\nonumber\\
P_{n}(t+1)&=&P_{n}(t).\nonumber
\end{eqnarray}
In these equations $\beta_n\equiv\Re\left[ a_n^*b_n\right]$ and $\Re(z)$ is the real part of $z$.
These terms take into account quantum coherence effects and are responsible for the essential
differences between the classical and quantum random walks. Under the assumption of a complete
decay of correlations their contribution is negligible and a classical description emerges  \cite{hadamard,master}.

At a given time and site only one among eqs.~(\ref{Pev}) applies, depending on the number
of broken links for the site. A statistical description can be obtained by combining them into
a single evolution equation with the appropriate statistical weights.
In the model described in Section \ref{sec:blinks}, the probability that a given site has no adjacent broken links is $\Pi_0=(1-p)^2$, that it has a left or right broken link is $\Pi_1=p(1-p)$ and that it is isolated is $\Pi_2=p^2$. These probabilities are appropriate statistical weights since they satisfy $\Pi_0+2\Pi_1+\Pi_2 = 1$. The resulting classical evolution equation for $P_n(t)$ is,
\begin{equation}
P_n(t+1)=pP_n(t) + \frac12 (1-p)\left[P_{n+1}(t)+P_{n-1}(t)\right]. \label{clas}
\end{equation}
This evolution describes a genuine diffusive process with a diffusion coefficient
\begin{equation}
D_{cl}=\frac 12 (1-p).\label{Dcbl}
\end{equation}
This can be easily seen by changing to continuous time and position variables $(t,x)$, in terms of which eq.~(\ref{clas}) becomes the diffusion equation
\begin{equation}
\frac{\partial P}{\partial t}=\frac 12 (1-p)\frac{\partial^2 P}{\partial x^2}.\label{difeq}
\end{equation}

Thus the classical version of the broken link model, with complete correlation decay, implies a diffusion coefficient which starts at the unbiased random walker value $1/2$ for $p=0$ and decreases as the frequency of broken links increases.  The fact that this classical diffusion coefficient does not scale as $(1-p)/p$ as in the quantum random walk with broken links, is due to persistent quantum correlations. As a result of these persistent quantum correlations the diffusion coefficient of the decoherent quantum walk, eq.~(\ref{Dbl}), is always larger than the corresponding classical one, eq.~(\ref{Dcbl}). This characteristic of the quantum walk has also been mentioned by other authors which have considered other sources of decoherence \cite{Brun,Shapira} and is also apparent in the case of repeated measurements considered in this work. 

\section{Brownian motion}
\label{sec:Brownian}

\begin{figure}[tbp]
\begin{center}
\includegraphics[scale=0.45]{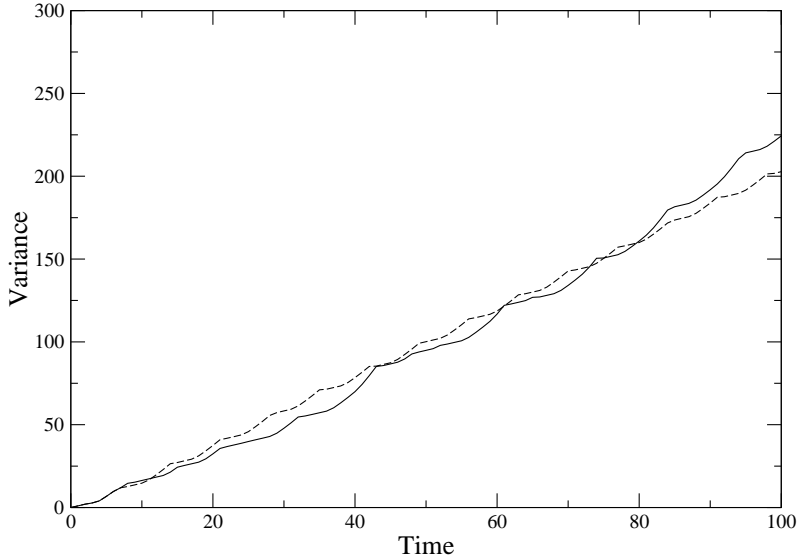}
\end{center}
\caption{ Time evolution of the average variance of the
quantum walker (full line) with random time intervals between measurements.
These intervals are uniformly distributed in [1,10]. The dashed line corresponds to the periodic measurement case with $T=7$}
\label{fig:dia3}
\end{figure}

The periodic measurement model discussed in Section~\ref{sec:medidas} can be easily generalized
to the case where the time intervals between consecutive measurements are randomly distributed.
In this case the diffusion coefficient is given by
\begin{equation}
D_{rm}=\frac{C\overline{T^{2}}}{2\overline{T}},  \label{ran}
\end{equation}
where $\overline{T}$ $(\overline{T^{2}})$ is the average of the (squared) time intervals between
measurements. The evolution of the average variance when the distribution of time intervals
between measurements is uniform in $[1,10]$ is shown in Fig.~\ref{fig:dia3}.
In this case, $\overline{T}=5.5$, $\overline{T^{2}}=38.5$ and the diffusion coefficient from
eq.~(\ref{ran}) agrees with that of the periodic case, eq.~(\ref{Drm}), for $T=7$.

This diffusion with random intervals between measurements is analogous to the
one found in the case of simple classical Brownian motion, which describes a heavy particle
immersed in a fluid of light molecules which collide with it randomly.
The collisions may be associated to the quantum measurement processes, while the
free motion between collisions plays the role of the unitary quantum evolution between
measurements.

To explore further this analogy, let us consider the Langevin equation for the
one-dimensional motion of a Brownian particle \cite{Reif}
\begin{equation}
\frac{dv}{dt} + \gamma v = f(t)  \label{brownian}
\end{equation}
where $v$ is the velocity of the particle, $\gamma$ a
viscosity coefficient and $f(t)$ the impulsive force per unit mass due to the
random collisions with the surrounding molecules. It is assumed that
$f(t)$ has a zero average and that its correlations satisfy
the Fluctuation-Dissipation theorem.
Under conditions of thermal equilibrium with the surrounding fluid and if initially $\overline{x}=\overline{x^2}=0$, the time evolution of the variance
for the position of the Brownian particle is
\begin{equation}
\sigma^2 = \frac{2C}{\gamma}
\left[t-\gamma^{-1}\left( 1-e^{-\gamma t}\right) \right],\label{browniano}
\end{equation}
where $C$ is a constant, fixed by the initial conditions, which is related to the temperature in the Brownian motion model. We now consider how this model applies to the decoherent quantum walks described in this work.

The inverse of the dissipation coefficient, $\gamma^{-1}$, defines a characteristic time in which the variance has a transition between a quadratic and a linear increase.
For short times there is a small chance of a collision and the variance
increases quadratically
\begin{equation}
\sigma^2\simeq Ct^2 \qquad (t\ll \gamma^{-1}) \label{var3}
\end{equation}
This expression is analogous to eq.~(\ref{sigmaq}), obtained in Section \ref{sec:medidas} for the periodic measurement case.

For long times, there are many collisions, the dissipative effects become dominant  and the variance increases linearly with time,
\begin{equation}
\sigma^2\simeq\frac{2C}{\gamma}t \qquad (t\gg \gamma^{-1})\label{superior}
\end{equation}
since the particle is slowed down and suffers a diffusive process with diffusion coefficient
\begin{equation}
D=\frac{C}{\gamma}.  \label{difusion}
\end{equation}
This can be compared to the quantum random walker at long times,
\textit{i.e.} after many position and chirality measurements have been performed.
From eq.~(\ref{Drm}) we obtain the relation between the period $T$ of the measurements for the quantum walker of Section \ref{sec:medidas} and the characteristic time of Brownian motion as
\begin{equation}
\gamma=\frac{2}{T}.  \label{period}
\end{equation}
Thus, repeated measurements can be viewed as a dissipative
process and the quantum random walker with decoherence brought about by these measurements
can be described in terms of a fully classical model. This suggests that the particular mechanism by which the decoherence is introduced in the dynamics does not seem to be important as long as the
proper time scale is considered.

\begin{figure}
\includegraphics[scale=0.8,angle=-90]{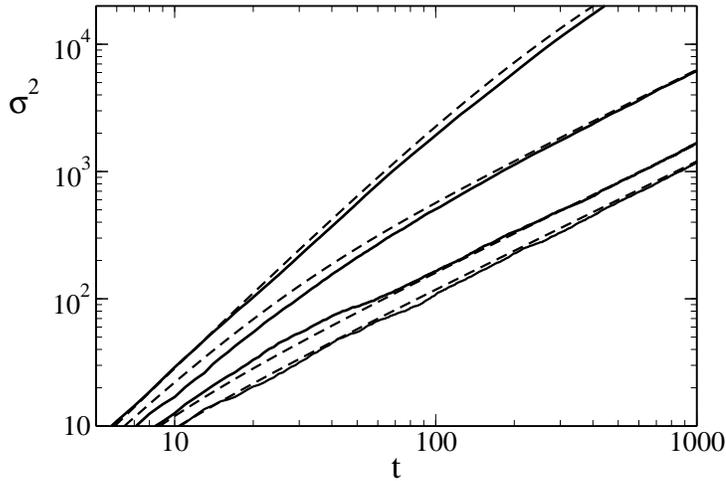}
\caption{Comparison (log--log scale) of the time evolution of the average
variance of the quantum walker with broken links (full lines) and of the
equivalent Brownian particle (dashed lines), as given by eq.~(\ref{browniano}) with
$C=0.293$ and $\gamma$ from eq.~(\ref{gblink}).
The frequencies $p$ of broken links are (from top to bottom) $p=0.01, 0.10, 0.30$ and $0.40$ respectively.}
\label{fig:langevin}
\end{figure}

In fact, if decoherence is introduced through broken links, as described in
Section~\ref{sec:blinks}, the Brownian motion model also applies and both
parameters of the Brownian model, $C$ and $\gamma$, can be read out from the 
numerical results. For early times, $t\ll t_c$, the
variance of the quantum walker increases quadratically because the chance of
decoherent events under the wave function, proportional to $pt$, is initially very low.
For the initial conditions used in this work, we have  $C=0.293$.
The open links under the wavefunction act as dissipative events, so the dissipation 
rate $\gamma$ is related to the frequency of broken links $p$. A comparison of the 
diffusion coefficients, eqs.~(\ref{difusion}) and~(\ref{Dbl}), implies the relation
\begin{equation}
\gamma = 0.73\,\frac{p}{1-p}.\label{gblink}
\end{equation}

As shown in Fig.~\ref{fig:langevin}, this simple stochastic model describes 
well the time evolution of the variance over a broad range of $p$ values, 
including the smooth transition that takes place between a quadratic and a 
linear increase regime in the variance.

\section{Conclusions}
\label{sec:conclusions}

Two instances of decoherent quantum walks have been considered. In one of them, decoherence is introduced through frequent measurements of position and chirality. In the other, decoherence results from randomly breaking a few links in the line. Through comparison of the results obtained with these two models and their classical counterparts, we have drawn some interesting conclusions about the diffusion rates and the decay of correlations in decoherent quantum walks.

In the case of the quantum walk on the line with periodic position and chirality measurements, it has been shown that the evolution, in a timescale involving many measurement events, can be described as a Markovian process and the probability distribution satisfies a Master equation. The associated diffusion coefficient has been analytically obtained and it varies inversely with the frequency of measurements. This results from the fact that the measurements
introduce decoherence in the quantum dynamics, partially breaking correlations and providing a route to a classical-like behavior. The periodic measurement has been extended to random time intervals between consecutive measurements and in this case the diffusion
coefficient depends on a characteristic time given by the ratio of the second to the first moment
of the time interval distribution. A comparison with the well known Markovian
process of Brownian motion allowed us to relate the average time between measurements with the characteristic time for Brownian motion, $\gamma^{-1}$.
Furthermore, the initial condition of the quantum random walker determines a constant $C$ which is  directly related to the temperature in the Brownian motion model.

When the decoherence is introduced by randomly breaking a some links in the line, a crossover from a coherent regime with the characteristic quadratic increase of the variance to a
diffusive regime with a linear increase of the variance results. We find numerically that in this case the diffusion coefficient scales with the link breaking probability $p$ as $(1-p)/p$ and that coherence effects are limited to a timescale $\sim 1/p$. The simple Brownian motion model adequately describes the time evolution of the variance, including the transition from quadratic to linear growth, for a broad range of decoherence rates.

These results suggest that in noisy quantum systems the coherence of the quantum evolution persists below a characteristic time of order $1/f$ where $f$ is the frequency of single decoherent events. After this characteristic time a quantum diffusion regime results in which the quantum diffusion rate is faster by a factor $1/f$ than the classical rate. This is due to quantum correlations present between consecutive decoherent events.
For low decoherence rates, even at long times, the decoherent quantum evolution diffuses faster than the corresponding classical evolution. On the other hand, at high decoherence rates, the role of quantum coherences is minimized and the quantum diffusion approaches the classical diffusion rate.

Acknowledgements:\\
G.A. and R.D. thank H. Pastawski for useful discussions.
We acknowledge support of \textit{PEDECIBA} and \textit{CONICYT} (Uruguay). R.D.
acknowledges partial financial support from the Brazilian Research Council
(CNPq). A.R, G.A. and R.D. acknowledge financial support from the
\textit{Brazilian Millennium Institute for Quantum Information}--CNPq.

\end{document}